\documentclass[12pt, a4paper]{openjournal}

\usepackage{physics}
\usepackage[english]{babel}
\usepackage{csquotes}
\usepackage{xcolor}
\usepackage{graphicx}
\usepackage{color}
\usepackage{tikz,pgf}
\usepackage{amsmath,amssymb,theorem}
\usepackage{url}
\usepackage{listings}
\usepackage{multirow}
\usepackage{booktabs}
\usepackage{xspace}
\usepackage{cancel}
\usepackage[printonlyused,withpage]{acronym}
\usepackage{rotating}
\usepackage{array}
\usepackage{stmaryrd}
\usepackage[htt]{hyphenat}
\usepackage[toc,page]{appendix}
\usepackage{tensor}
\usepackage[breaklinks,colorlinks]{hyperref}
\hypersetup{
	colorlinks, 
	linktoc=all, 
	linkcolor=blue,  
	citecolor=hyptxt,
	urlcolor=hyptxt}
\hypersetup{
	citebordercolor=,
	filebordercolor=green,
	linkbordercolor=blue
}
\definecolor{hyptxt}{rgb}{0.7, 0.4, 0.9}
\usepackage{scrhack} 

\usepackage{newtxtext} 

\newlength{\myleftlen}

\begin{document}


\title{Black Hole Shadow Drift and Photon Ring Frequency Drift}

\author{Emmanuel Frion}
\email{emmanuel.frion@helsinki.fi}
\affiliation{University of Helsinki, Department of Physics and Helsinki Institute of Physics,\\ P.O. Box 64, FIN-00014 University of Helsinki, Finland}

\author{Leonardo Giani}
\email{uqlgiani@uq.edu.au}
\affiliation{The University of Queensland, School of Mathematics and Physics,\\ QLD 4072, Australia}

\author{Tays Miranda}
\email{tays.miranda@helsinki.fi}
\affiliation{University of Helsinki, Department of Physics and Helsinki Institute of Physics,\\ P.O. Box 64, FIN-00014 University of Helsinki, Finland\\Department of Physics, University of Jyväskylä, P.O. Box 35, FIN-40014  Finland}

\begin{abstract}
	\noindent
	 The apparent angular size of the shadow of a black hole in an expanding Universe is redshift-dependent. Since cosmological redshifts change with time - known as the redshift drift - all redshift-dependent quantities acquire a time dependence, and \textit{a fortiori} so do black hole shadows. We find a mathematical description of the black hole shadow drift and show that the amplitude of this effect is of order $10^{-16}$ per day for M87$^{\star}$. While this effect is small, we argue that its non-detection can be used to constrain the accretion rate around supermassive black holes, as well as a novel probe of the equivalence principle. If general relativity is assumed, we infer from the data obtained by the Event Horizon Telescope for M87$^{\star}$ a maximum accretion rate of $|\dot{M}/{M}| \leq 10^5 M_{\odot}$ per year. On the other hand, in the case of an effective gravitation coupling, we derive a constraint of $|\dot{G}/G| \leq 10^{-3}-10^{-4}$ per year. The effect of redshift drift on the visibility amplitude and frequency of the universal interferometric signatures of photon rings is also discussed, which we show to be very similar to the shadow drift. This is of particular interest for future experiments involving spectroscopic and interferometric techniques, which could make  observations of photon rings and their frequency drifts viable.
\end{abstract}

\maketitle

\section{Introduction}

Redshift is an omnipresent Doppler-effect-related quantity, used in cosmology to build meaningful spatial and temporal distances. In 1962, Sandage and McVittie \cite{sandage1962change,mcvittie1962appendix} showed that the redshift is in fact a dynamical quantity, with a time derivative $\dot{z}$, referred to as redshift drift. A redshift drift driven by the expansion of the Universe is called a cosmological drift\footnote{Other contributions from peculiar velocities and inhomogeneities exist, and are discussed in \cite{linder2010constraining}.}, with the interesting consequence of turning  redshift-dependent quantities into  time-dependent ones. The cosmological drift depends on the Hubble rate $H(z)$, and is expected to be very small at low redshift $z$, with a redshift change of the order of $10^{-10}$ per year. However, it applies to every object in the Universe, which makes all of them  possible probes of the drift, and collaborations such as the Extremely Large Telescope \cite{Liske:2008ph}, the Square-Kilometer Array \cite{Klockner:2015rqa} or the Vera C.~Rubin Observatory \cite{LSST:2017ags} will provide means of its detection. Optimistic estimates show that these facilities could  reach a precision of $10^{-10}$, for example with monitoring programs of 1000 hours of exposure with a 40-meter telescope \cite{Kim:2014uha}. For this reason, cosmology with redshift drifts is an active field of research  \cite{TeppaPannia:2013lbl,Mishra:2014vga,Martins:2016bbi,Pandolfi:2014nfa,Alves:2019hrg,Piattella:2017uat,Melia:2016bnb,TeppaPannia:2018ale,Bolejko:2019vni,Martinelli:2012vq,Quartin:2009xr,Giani:2020fpz,Codur:2021wrt}, which we advance further with this work, in which we investigate how the cosmological drift would affect the image from black hole shadows and interferometric signatures from black holes' photon rings. 

Black holes were first predicted over a century ago, yet the first direct image of a black hole was reconstructed only very recently by the Event Horizon Telescope (EHT) collaboration \cite{Akiyama:2019bqs,Akiyama:2019brx,Akiyama:2019cqa,Akiyama:2019eap,Akiyama:2019fyp,Akiyama:2019sww,Akiyama:2021qum}. The direct detection of the shadow of M87$^{\star}$, the supermassive black hole at the centre of the galaxy Messier 87, suggests we will soon be able to obtain black hole parameters such as their mass, spin and charge in a systematic way. Black hole shadows, which are essentially dark images of the black holes' event horizon, and photons rings, which in contrast are bright images formed by photons orbiting the black holes, are observables that can be used to infer these parameters \cite{Bardeen:1973tla,Bambi:2019tjh,Allahyari:2019jqz,Chowdhuri:2020ipb,Hadar:2020fda,Himwich:2020msm,Ghosh:2020spb,Afrin:2021imp,Broderick:2021ohx,Roelofs:2021wdi}, and are particularly interesting to precisely constrain gravity in the strong-field regime \cite{Vagnozzi:2019apd,Khodadi:2020gns,Khodadi:2020jij,Gralla:2020srx,Psaltis:2020lvx,Khodadi:2021gbc,Li:2021mzq,Aratore:2021usi}. In fact, black hole shadows could potentially be used as standard rulers to infer the present day value of the Hubble parameter $H_0$ once that the mass of the black hole has been inferred independently \cite{Tsupko:2019pzg,Qi:2019zdk}. These aspects (and more) about black hole shadows are covered in the recent review \cite{Perlick:2021aok}. 

There exist some issues associated to the direct observations of supermassive black holes, though, as discussed by Vagnozzi, Bambi and Visinelli in \cite{Vagnozzi:2020quf}. The determination of black holes' masses, for example, usually differ when considering stellar dynamics or gas dynamics measurements, and this discrepancy directly alters the precision associated to the measurement of the black holes' distances. The apparent angular diameter of a black hole shadow depends on both the black hole's mass and distance from the observer, so precision cosmology using black holes could be hampered by this lack of precision. Finding more sources we could retrieve information from would partly compensate this issue, and enhancing the EHT sensitivity is the next goal of the collaboration, though reaching a resolution ten times better (\textit{i.e.} of about 0.1 $\mu as$) requires futuristic facilities such as the implementation of lunar telescopes. To make matters even more difficult, only from low-redshift black holes would we extract information relatively easily, as high-redshift black holes 1) are much more difficult to observe due to the loss of surface brightness with redshift, and 2) their formation history is poorly known.

Even within the assumption that we can resolve the shadow of many black holes, the surrounding matter falling onto a black hole brings two additional challenges. The first of them is that the light emitted during accretion, clearly visible in the image of M87$^{\star}$ resolved by the EHT, largely dominates the light shone by photon rings. This image was obtained using a baseline with roughly the same size of the diameter of the Earth. To distinguish the luminosity emitted by the rings, much longer baselines are required \cite{Johnson:2019ljv,Gralla:2020nwp}. While the luminosity flux of photon rings is exponentially decreasing with each orbit, the signal is dominated by periodic universal interferometric signatures for baselines greater than approximately 20 G$\lambda$ \footnote{Baselines are often given in units of wavelengths. This is because the telescope resolution (in radians) is obtained through the relation $R=\lambda/u$, with $u$ the baseline. Therefore, when $u$ is proportional to the wavelength, the resolution is simply $R=1/u$.} (see figure 4 of \cite{Johnson:2019ljv}), with the period inversely proportional to the diameter of the ring. The second challenge is that accreting matter changes the apparent angular diameter with time. Accretion processes are model-dependent, and strongly depend on the shape and inclination of the accretion disk, as well as on the accretion flow. In this work, we first show that cosmological expansion affects both the angular size of black hole shadows and the periodic signal in the same fashion. After this, we discuss the possibility of using the non-detection of a black hole shadow drift as a method to constrain accretion rates in an almost model-independent way. We also show that it can be used in modified gravity models, as another probe of the equivalence principle.

\section{Black hole shadow drift}

\subsection{Schwarzschild black hole in an expanding universe} \label{sec_schwarz}

The McVittie metric \cite{McVittie:1933zz} is a hybrid solution recovering the Schwarzschild metric near the black hole and the flat Friedmann-Lemaître-Robertson-Walker (FLRW)  far away from it. Therefore, when the metric is well-defined, it describes a spherically symmetric black hole in an expanding Universe. Its line element is usually written as
\begin{align}
	\textup{d} s^2=-\left(\frac{1-\mu}{1+\mu}\right)^2 c^2 \textup{d} t^2 + (1+\mu)^4 \, a^2(t) \, \left(\textup{d} l^2+l^2\textup{d} \Omega^2  \right) \;,
\end{align}
with
\begin{align}
	\mu := \frac{m}{2a(t) l} \;.
\end{align}
In this definition, $l$ is the radial comoving distance, $a(t)$ is the scale factor characterising the expansion of the Universe, and $m=GM/c^2$, with $M$ the mass of the black hole. For an observer near the black hole, the scale factor is nearly constant and evaluated at $t=t_0$, which allows to recover the Schwarzschild metric with the change of variable
\begin{align}
	R = r \left(1+\frac{2m}{r}\right)^2 \;,
\end{align}
and the definition $r:=a(t_0)\, l$. The position of the observer at time $t_0$ is denoted $R_O$ in these coordinates, and $t_0$ corresponds to the present time.

An approximate solution for black hole shadows in a McVittie spacetime has been found as a composite solution by employing the technique of matching solutions for the Einstein equations \cite{Tsupko:2019mfo}:
\begin{align}
	\alpha_{\textup{appr}}(R_O) = \alpha_{\textup{schw}}(R_O) + \alpha_{\textup{cosm}}(R_O) - \alpha_{\textup{overlap}}(R_O) \;,
\end{align}
where the three pieces on the right side of the equation are respectively the angular diameter of the shadow when observed from near the Schwarzschild black hole, in an intermediate region far from the black hole but in which the cosmological expansion is  negligible, and at cosmological scales. These are defined as:
\begin{align}
	\alpha_{\textup{schw}}(R_O) &=\left\{ \begin{aligned}
		\pi -\arcsin({3\sqrt{3}m\sqrt{1-2m/R_O}/R_O}) 
		\quad &\text{for}\, 2m\leq R_O \leq 3m \\ \arcsin({3\sqrt{3}m\sqrt{1-2m/R_O}/R_O})\quad &\text{for}\,  R_O \geq 3m \\
	\end{aligned} \right.  \\
	\alpha_{\textup{overlap}}(R_O) &= \frac{3\sqrt{3}m}{R_O} \hspace{6.05cm} \text{for}\, m\ll R_{int} \ll c/H_0 \\
	\alpha_{\textup{cosm}}(R_O) &= \frac{3\sqrt{3}m}{D_A(z)} \hspace{6.05cm} \textup{otherwise} \;. \label{cosmosol}
\end{align}
The cosmological solution depends on the redshift $z$ \cite{Bisnovatyi-Kogan:2018vxl}, and, in a flat universe, the apparent size for an observer at cosmological distances depends on the angular diameter distance $D_A(z)$ defined by
\begin{align}
	\label{angular_distance}
	D_A(z):=\frac{c}{1+z} \int_0^{z} \frac{d\Tilde{z}}{H(\Tilde{z})}=\frac{1}{1+z}\chi\;,
\end{align}
with $\chi$ the comoving distance to the black hole.

The double inequality in the overlap solution indicates the region in which the approximate solution is valid. In other words, if we can find a distance $R_{int}$ in which both the expansion of the Universe and the gravity field produced by the black hole are negligible, the approximation solution holds. This should always be the case in practice, since the size of a black hole is always several orders of magnitude smaller than the scale at which the expansion is important (which could be supposed to be the scale of galaxy clusters).


\subsection{Shadow drift}

By construction, only the cosmological term \eqref{cosmosol} of the composite solution is redshift-dependent. For an observer sufficiently far away from the black hole, the overlap solution and the Schwarzschild solution cancel. Therefore, any drift in the shadow angular radius is fully contained into the relation (omitting $R_O$ for conciseness)
\begin{align}
	\frac{\dot{\alpha}_{\textup{appr}}}{\alpha_{\textup{appr}}} = \frac{\dot{\alpha}_{\textup{cosm}}}{\alpha_{\textup{cosm}}} \;.
\end{align}
The time derivative of $\alpha_{\textup{cosm}}$,
\begin{align}
	\dot{\alpha}_{\textup{cosm}} = -3\sqrt{3} m \frac{\dot{z}}{D_A^2(z)} \frac{\textup{d}D_A}{\textup{d}z} = -\alpha_{\textup{cosm}} \frac{\dot{z}}{D_A(z)}\frac{\textup{d}D_A}{\textup{d}z}\;,
\end{align}
explicitly depends on the variation of the angular diameter distance \eqref{angular_distance} with redshift and on the redshift drift $\dot{z}$. The first quantity is
\begin{align}
	\label{angdistder}
	\frac{\textup{d}D_A}{\textup{d}z} = -\frac{D_A(z)}{1+z} \;.
\end{align}
Recalling the redshift drift in a FLRW background is \cite{sandage1962change,mcvittie1962appendix,linder1997first,Piattella:2015xga}
\begin{align}
	\label{redshift_drift}
	\frac{\textup{d}z}{\textup{d}t} = H_0 (1+z) -H(z)\;,
\end{align}
we find the \textit{shadow drift}
\begin{align}
	\label{shadow_drift}
	\frac{\dot{\alpha}}{\alpha} =  H_0 -\frac{H(z)}{1+z} \;,
\end{align}
in which we have dropped the subscript in the angular radius for conciseness. 

\begin{figure}
	\centering
	\begin{minipage}[c]{\textwidth}
		\centering
		\includegraphics[scale=0.35]{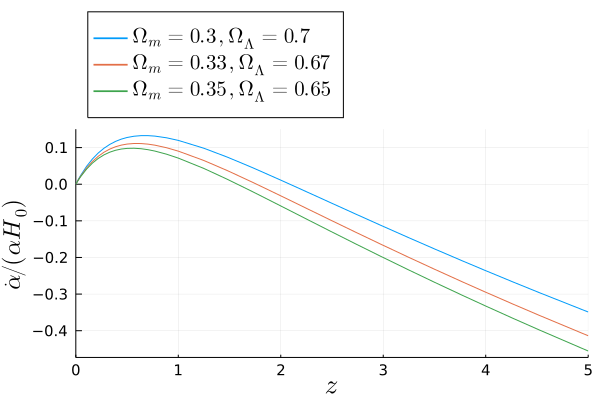}
		\includegraphics[scale=0.35]{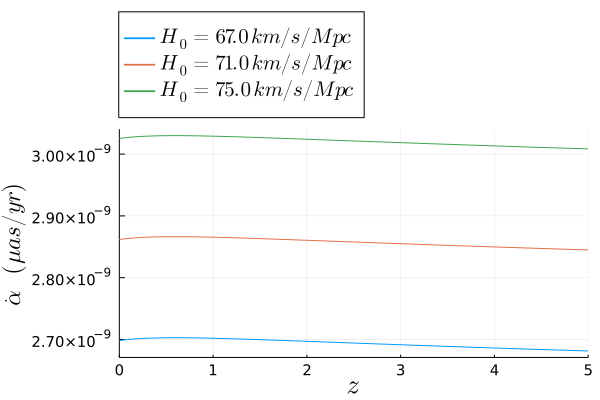}
		\caption{Left: Normalised shadow drift as a function of redshift for three sets of matter and dark energy dimensionless densities. At low redshift, the maximum normalised shadow drift is about $0.1$ for the three sets, which results in a shadow drift $\dot{\alpha}/\alpha$ of $10^{-14}$ day$^{-1}$. Right: Variation of the apparent angular diameter for three different values of the present Hubble rate. We used density values of $\Omega_m=0.33$ and $\Omega_{\Lambda}=0.67$, and assumed an apparent angular diameter $d=40 \mu as$.\vspace{1cm}}
		\label{fig_shadow_drift}
	\end{minipage}
\end{figure}

In a flat $\Lambda$CDM universe, the Hubble parameter is related to the matter density $\Omega_{m0}$, the radiation density $\Omega_{r0}$ and the dark energy density $\Omega_{\Lambda}$ through
\begin{align}\label{H(z)LCDM}
	H(z) = H_0 \sqrt{\Omega_{m0}(1+z)^3+\Omega_{r0}(1+z)^4+\Omega_{\Lambda}} \;.
\end{align} 
In Fig.~\ref{fig_shadow_drift}, the left figure shows the shadow drift normalised with $H_0$ for three sets of dimensionless densities, assuming $\Omega_{r0}=0$. For each mode, drift effects get much higher with redshift, but the luminosity drops quickly, which makes high-redshift black hole shadows impractical targets in real situations for the current status of observations. Indeed, the observed flux of a source located at redshift $z$ decreases as $(1+z)^2$, while the surface brightness as $(1+z)^4$, so a source at $z=1$ would see its surface brightness reduced by a factor 16  \cite{Vagnozzi:2020quf}. We note that there is a small bump at low redshift around $z=0.5$ where the drift is potentially easier to observe. In the right figure, we display the variation of the angular apparent radius $\dot{\alpha}$. The plot shows that, for a given cosmology (\textit{i.e.} a given set of density parameters), it is possible to infer the value of the Hubble parameter today $H_0$ if the black hole mass is obtained independently. We have used the density values $\Omega_{m0}=0.33$, $\Omega_{r0}=0$ and $\Omega_{\Lambda}=0.67$ to obtain this figure.

\subsection[Shadow drift of M87*]{Shadow drift of M87$^{\star}$}

Let us work out as an example the shadow of the black hole at the centre of the galaxy Messier 87, M87$^{\star}$. The heliocentric radial velocity of M87$^{\star}$ is around 1284 km/s \cite{cappellari2011atlas3d}, with a peculiar velocity of the order of $\sim$ 10\% of the total velocity \cite{lisker2018active}, which we assume to be too small to have a noticeable impact on the shadow drift. We also assume that the contribution from the peculiar accelerations to be negligible, \textit{i.e.}, we suppose the peculiar velocity will not change noticeably. Furthermore, the estimated distance of  M87$^{\star}$ from us is roughly $z\sim 0.004$ \cite{cappellari2011atlas3d}, which ensures the distance of the observator $R_O$ is at cosmological scales, and that we can use the results of the previous section. At sufficiently low redshift ($z\ll1$), we can safely ignore the contribution from radiation in Eq. \eqref{H(z)LCDM}. We use the density values $\Omega_{m0}\simeq 0.31$, $\Omega_{\Lambda}\simeq 0.67$, and the Hubble rate today $H_0 = 67.8$ km/s/Mpc.

The shadow drift is then:
\begin{equation}
	\frac{\dot{\alpha}}{\alpha} \approx 2 H_0 \times 10^{-3}\approx 3.8 \times 10^{-16}/day \;.
\end{equation}
A tiny variation, which seems far away from the angular resolution the EHT can achieve from Earth.\footnote{It must be stressed that the EHT is currently capable of resolving objects of size $d\approx 25 \mu as$, even though through imaging techniques they were able to reach a sensitivity of the order of $\approx 1\mu as$ for M87$^*$ and SgrA$^*$ (see details in EHT IV \cite{Akiyama:2019bqs}).}  Indeed, even for the future Plateau de Bure--South Pole baseline, which should reach a resolution of 15 $\mu$as at 345 GHz, this seems like an infeasible task. An Earth--Moon baseline could potentially enhance the resolution by a factor 10, while an Earth-L2 Lagrange point baseline by a factor 100, but does not raise much hope for this detection. 

\section{Shadow drift as a probe of the equivalence principle and of accretion rates}

We argue in this section that the impossibility of detecting the cosmological shadow drift can be put into good use. Until now, we have considered that only the redshift depends on time, though it would be interesting to let $m$ be a time-dependent quantity as well in the cosmological solution \eqref{cosmosol}. In this case, the three quantities $G$, $M$ and $c$ could depend on time, but we will keep the speed of light constant in this work.  

\begin{figure}
	\centering
	\begin{minipage}[c]{\textwidth}
		\centering
		\includegraphics[scale=1.2]{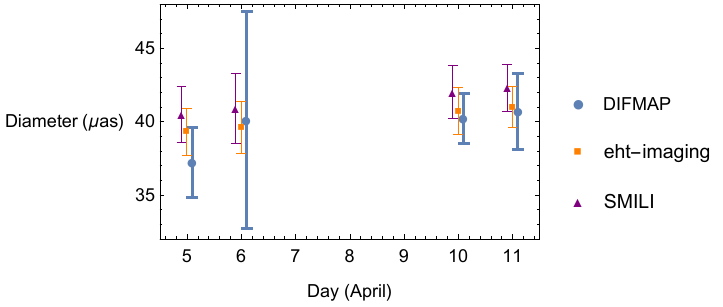}
		\caption{The observed diameter of the shadow of M87* measured by the EHT with three different pipelines, DIFMAP, eht-imaging and SMILI, from April 5 to April 11. Data taken from Table 7 of \cite{Akiyama:2019bqs}.}
		\label{diametervstime}
	\end{minipage}
\end{figure}

First, we assume a time-dependent gravitational coupling and a constant mass. Then the right-hand side of the shadow drift \eqref{shadow_drift} has an additional contribution
\begin{align}
	\label{equiv}
	\frac{\dot{\alpha}}{\alpha} =  H_0 -\frac{H(z)}{1+z}+\frac{\dot{G}}{G} \;.
\end{align}
Therefore, the non-observation of a drift in the angular shadow within the sensitivity of present experiments provides another probe of the equivalence principle, complementary to other works using black hole shadows to test the equivalence principle \cite{Li:2019lsm}. Using the mean values of M87$^{\star}$ shadow's angular diameter reported by the EHT collaboration, which we also plot in Fig. \ref{diametervstime},  during the period of observations from April 5 to April 11 (see Table 7 in EHT IV \cite{Akiyama:2019bqs}), we compute the shadow drift between these two dates for the three different pipelines (DIFMAP, eht-imaging and SMILI) used by the collaboration, summarised in Table \ref{tabledrift}. 

\begin{table}[h!]
	\centering
	\begin{tabular}{c||c|c|c|}
		&DIFMAP & eht-imaging & SMILI\\
		\hline
		$\dot{\alpha}/ \alpha$ ($10^{-2}$) & $1.6 \pm 1.7$ & $0.7 \pm 0.9$ & $0.7 \pm 0.9 $
	\end{tabular}  
	\caption{Estimated variation of the angular size of M87$^{\star}$'s shadow for the three pipelines used by the EHT collaboration. All values are given in day$^{-1}$.}
	\label{tabledrift}
\end{table}

We can readily see from the table that the non-observation over a period of 6 days implies that  $\dot{G}/G \lesssim  10^{-1}-10^{-2}$ day$^{-1}$. Extrapolating over a period of one year would improve the precision to roughly $10^{-3}-10^{-4}$ year$^{-1}$. While this precision is far inferior to other existing probes (the strongest constraint gives $\dot{G}/G \leq 10^{-13}-10^{-14}$ day$^{-1}$, see for example Table 1 from Ref.~\cite{Alestas:2021nmi}), we note that the shadow drift only depends on the cosmological model. Therefore, it is an almost model-independent probe of the equivalence principle similar to the one found in \cite{Giani:2020fpz} in the case of strong lensing.

On the other hand, it must be stressed that a variation of the effective gravitational coupling $G$ is degenerate with a variation of the black hole mass $M$. Modeling the evolution of a black hole mass through accretion is a cumbersome task, involving many astrophysical processes which ultimately depends on the black hole environment, as discussed for example in Ref.~\cite{Li:2012ts}. Consequently, in order to properly employ drift observations of rings and shadows to test the equivalence principle, a reliable accretion model is required. However, assuming that General Relativity is the correct theory of gravitation, for which $G$ is constant, the very same argument shows that the non-observation of a shadow drift can be used to constrain accretion rates $\dot{M}/M$, since we would now have
\begin{align}
	\frac{\dot{\alpha}}{\alpha} =  H_0 -\frac{H(z)}{1+z}+\frac{\dot{M}}{M} \;.
\end{align}
For M87$^{\star}$, which has a mass of roughly $6.5 \times 10^9 M_{\odot}$ as measured by the EHT collaboration \cite{Akiyama:2019eap}, an absence of shadow drift of order $\dot{\alpha}/{\alpha} \leq 10^{-4}$ year$^{-1}$ implies that the black hole mass has changed less than $10^5 M_{\odot}$ over a year. \footnote{Note that superradiance could potentially change the apparent angular diameter of the shadow \cite{Roy:2019esk,Creci:2020mfg}, and drift effects could have interesting applications for these models as well.}

One way to distinguish between a time dependence of the mass and a time dependence of $G$ is to look at the variation sign of $\dot{\alpha}/\alpha$. Since evaporation is a very slow process, we expect that only accretion would change the mass of the black hole, and therefore $\dot{M}/M >0$. This implies that if we observe $\dot{\alpha}/\alpha <0$, the variation of the angular diameter of the shadow can be attributed to a variation in the gravitational coupling, and to the cosmological drift (even if the latter is very small as we discussed in the previous section). To have $\dot{M}/M >0$, the black hole needs to accrete matter from its surroundings. The maximum efficiency of this process is given by the Eddington rate (see \textit{e.g.} \cite{Brito:2014wla})
\begin{align}
	\dot{M}_{Edd} := 0.02 f_{\textrm{Edd}} \frac{M}{10^{6} M_{\odot}} M_{\odot} \textup{yr}^{-1} \;,
\end{align}
where $f_{\textrm{Edd}}$ is the Eddington ratio typically ranging between $10^{-2}$ and $1$ \cite{Barausse:2014tra}. Note that this formula assumes a radiative efficiency $\eta \approx 0.1$. For M87, the Eddington ratio is expected to be at most $f_{\textrm{Edd}}=0.03$ \cite{Forman:2017kpv}, resulting in $\dot{M}_{\textrm{Edd}}/M \simeq 6 \times 10^{-10} M_{\odot} \textup{yr}^{-1}$, or, equivalently, in $\dot{M}_{\textrm{Edd}} = 4 M_{\odot} \textup{yr}^{-1}$. The latter estimate implies that the strongest constraint on $\dot{G}/G$ which could be obtained from M87$^{*}$ is of order $10^{-10} $ yr$^{-1}$.
Alternatively, drifts observations could be used to constrain models of BH accretion, for example to obtain upper bound on the Eddington rate $f_{\textrm{Edd}}$.

\section{Ring visibility amplitude and frequency drifts}

We investigate in this section a complementary possibility of observing drift effects with black holes using photon rings, which should be observable for interferometers with baselines $u$ at least greater than 20 $G\lambda$ (or, equivalently, for a resolution $1/u$ better than $10 \, \mu as$). For such large baselines, the complex visibility $V(u)$ of a Kerr photon ring can be approximated by a damped oscillating function with period $\Delta u=2/d$, with $d$ the ring diameter \cite{Johnson:2019ljv}. As argued by the authors of \cite{Johnson:2019ljv}, all rings are almost circular, independently of the black hole spin and inclination, so we adopt in this section a perfectly circular photon ring model to study the effect of cosmological expansion on the ring diameter. Since the photon ring properties in Schwarzschild do not differ appreciably from Kerr black hole photon rings \cite{Gralla:2019xty}, we assume the conclusions drawn by Johnson et al in \cite{Johnson:2019ljv} hold for a Schwarzschild black hole. For simplicity, we suppose further that all rings are infinitesimally thin and uniform. In this case, the complex visibility is given by a Bessel function of the first kind, $V(u)=J_0(\pi du)$, which reduces to roughly $(du)^{-1/2} \cos(\pi du)$ for large arguments.

\begin{figure}
	\centering
	\includegraphics[scale=0.4]{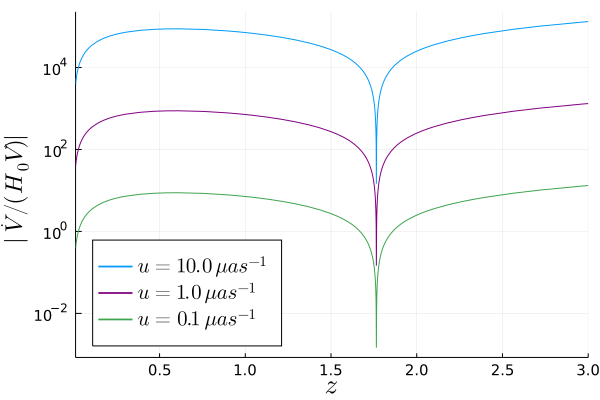}
	\caption{Variation of the visibility amplitude for a photon ring with apparent angular diameter $d=40 \,\mu as$. The variation is a function of the redshift and drawn for three baselines ($u=10,1,$ and $0.1 \,\mu as^{-1}$, in blue, purple and green, respectively).  For a redshift $z=0.004$ (M87$^{\star}$), the respective variations of the normalised amplitude $|\dot{V}/(H_0 \,V)|$ are of about $1.6\times 10^{3}$, $16.0$ and $0.16$ for each baseline, resulting in a change of the amplitude $|\dot{V}/V|$ of about a factor $3 \times 10^{-11}$, $3\times 10^{-13}$ and $3\times 10^{-15}$ per day.}
	\label{visibility_drift}
\end{figure}

Within the above hypotheses, we infer that the ring diameter $d$ should not change in time near the black hole. However, as we have shown in the previous sections,  the expansion of the Universe makes $d$ a time-dependent quantity, with a redshift drift similar to Eq.~\eqref{shadow_drift}. Therefore, the ring diameter acquires in this picture a time dependence
\begin{equation}
	\label{vis_drift}
	\frac{ \dot{d}}{d} = H_0 -\frac{H(z)}{1+z} \;,
\end{equation}
and the relative variation of the visibility amplitude varies as 
\begin{align}
	\frac{\dot{V}}{V} &= - \pi u \dot{d} \, \frac{J_1(\pi d u)}{J_0(\pi d u)} \nonumber \\
	&\simeq - \frac{1}{2} (\pi u d)^2 \frac{\dot{d}}{d} \nonumber \\
	&\simeq - \frac{1}{2} \left(\pi u d\right)^2 \left(H_0 -\frac{H(z)}{1+z}\right)\;,
\end{align}
where we have used the Bessel expansion for large argument $J_1(x)/J_0(x) \simeq x/2$ in the second line. This expansion is justified since the first photon ring should be resolved for large baselines, which implies $du\gg1$. The visibility variation interestingly depends on $u^2$, meaning that a greater baseline should probe a greater variation, as shown in Fig.~\ref{visibility_drift}.
The effect is once again very small, with a best case of a variation about $10^{-6}$ per year for $z\simeq 0.5$ using a baseline $u=10 \, \mu as^{-1}$, which would be barely attainable with a Moon-Earth array (see Fig.~5 in \cite{Johnson:2019ljv}). It is worthwhile though to mention that spectroscopic measurements are usually more precise than optical ones. It is expected that forthcoming surveys in the next few years will be able to detect a shift in the spectral lines of the order of $10^{-9}-10^{-10}$ within an observational window of $\sim 10 yr$, see for example Ref.~\cite{Kim:2014uha}. Therefore, if a similar precision can be reached for interferometric observations of photon rings, the task of measuring their drift does not seem hopeless. 

A second interesting point is that the period also suffers from the redshift drift, with a relative change 
\begin{equation}
	\label{period_drift}
	\frac{\dot{\Delta u}}{\Delta u} = - \frac{ \dot{d}}{d} = -\left(H_0 -\frac{H(z)}{1+z}\right) \;,
\end{equation}
opposite to the drift in the ring diameter. Since the $2/d$ periodicity is unlikely to be contaminated by other sources of emission on great baselines, the period is a universal signature which can be used to detect a black hole photon ring via the observed visibility. As an immediate consequence following from Eq.~\eqref{period_drift}, the universal interferometric pattern should also change with time, along with the strength of the visibility amplitude.

\section{Conclusions and Discussion}

The main purpose of this work was to quantify the impact of cosmological drift on three important black hole physics observables: the black hole shadow's  apparent angular diameter, the visibility, and the frequency (or equivalently the period) of its photon rings. The apparent angular diameter of the shadow, for a spherically symmetric black hole, is determined by its total mass $M$ and the angular diameter distance $D_A$, which makes it a possible candidate standard ruler, see Refs.~\cite{Tsupko:2019pzg,Qi:2019zdk}. Similarly, the visibility amplitude and frequency inferred from interferometric universal signatures of photon rings are mostly determined by the ring diameter, which could potentially be a source of cosmological information, as discussed in Ref.~\cite{Johnson:2019ljv}.
The drifts of these three quantities are given by Eqs.~\eqref{shadow_drift}, \eqref{vis_drift} and \eqref{period_drift}, respectively. In Fig.\ref{fig_shadow_drift}, the shadow drift is reported in units of $H_0$ up to redshift $z=5$ assuming a fiducial $\Lambda$CDM cosmology with $\Omega_m = 0.33$ and $\Omega_\Lambda = 0.67$. At low redshift, $z \ll 1$, the maximum drift is of order $10^{-1} H_0$. For M87$^{\star}$, located at redshift $z \approx 4\times 10^{-3}$, this results in a shadow drift of the order of $10^{-16}$ day$^{-1}$,  which is beyond the angular resolution of the EHT and forthcoming experiments. Concerning the status of photon ring observations, current experiments do not have enough sensitivity to resolve individual rings. On the other hand, as discussed in Ref.~\cite{Johnson:2019ljv}, future Earth-Moon and Earth-L2 baselines will have enough precision to resolve the rings and allow for spectroscopic observations. In Fig.~\ref{visibility_drift}, the visibility amplitude drift is shown for three future experiments with baselines larger than the one currently in use with the EHT (which tops at 0.05 $\mu as^{-1}$).  The shift induced in the spectral lines by the Hubble flow is the most promising candidate to detect redshift drift effects and, as discussed in Ref.~\cite{Kim:2014uha}, can reach a precision of order $10^{-9}$ over a time span of a decade. If, in the future, such a precision can be reached for spectroscopic measurements of photon rings, we might very well be able to observe their visibility amplitudes and frequencies drifts.

It is interesting to note that the formula for the black hole shadow \eqref{shadow_drift} is very similar to the one obtained for the lens equation in the thin-lens approximation, being proportional to the product of the Newton constant and the object mass $G_N M$ and inversely proportional to the angular diameter distance. 
In the same fashion as what has been done for strong lensing observables \cite{Giani:2020fpz}, it is possible to translate non-observations of drift effects within  the sensitivity of current experiments into a constraint on the violation of the Equivalence Principle in theories featuring a time-dependent gravitational coupling $G_{eff}$.
We quantified the non-observations of a drift in the size of M87$^{\star}$'s shadow over a period of one year, within the current sensitivity of EHT, into a constraint on $\dot{G}_{eff}/G_{eff} \leq 10^{-4}$ yr$^{-1}$. This constraint, however, is obtained assuming that the mass of the black hole $M$ does not vary with time. On the other hand, if this is the case, variations of the  black hole shadow within the framework of General Relativity can be used to verify accretion models of black holes. Using the current sensitivity of EHT, the non-observations of a shadow drift over a period of one year gives a constraint on the variation of M87$^{\star}$'s mass of $\Delta M \leq 10^5 M_{\odot}$. 

While most of the effects we presented in this paper are beyond the sensitivity of current experiments, it is worthwhile to recognize and quantify the amount of cosmological information encoded within black hole shadows for future long-monitoring programs. Furthermore, these observations have the potential to become an important test of the equivalence principle in a strong gravity regime, thus complementing other existing probes. 

In this work, we considered variations in the shadow of a single, spherically symmetric Black hole embedded in a FLRW Universe. It would also be very interesting to study how the results we obtained change when we relax these assumptions. As was discussed in \cite{Li:2020drn}, the shadow of a Kerr black hole can be obtained in a similar way as for a McVittie black hole, which we used in this paper. The shadow obtained by these authors is quite similar to those in McVittie, with two different angular sizes due to the black hole spin. By reproducing the steps of our paper to this Kerr-de Sitter framework, we obtain the same qualitative results. However, the embedding of a Kerr black hole into a FLRW background is not unique, as the former breaks not only homogeneity of the spacetime (like in McVittie) but also isotropy, introducing dragging effects relative to the rotation axes which makes the results observer dependent.

To conclude, we mention that it would be very interesting to study the time dependence of the shadow(s) profile(s) for two or more interacting black holes \cite{Yumoto:2012kz}. Indeed, in this case one expects that the mass loss due to the production of gravitational waves (GW) should correspond to a shrinking of the shadow(s). However, it must be stressed that the GW signal during the inspiral phase depends on a particular combination of the black hole masses, \textit{i.e.} the chirp mass, whose relation with the diameter of the shadow(s) is not trivial. Furthermore, the GW response function is affected in a similar way by the cosmological expansion, by a time-dependent gravitational coupling and by a time-varying mass \cite{Yunes:2009bv}. As a result, if one tries to relate the GW signal with a drift of the shadow(s), these effects will all be degenerate in the final measurement.  We believe that these directions deserve further investigations, which we leave for future works.

\section*{Acknowledgments}

We are grateful to Tamara Davis, Oliver Piattella, Oleg Tsupko, and Sunny Vagnozzi for valuable comments and discussions.
EF thanks the Helsinki Institute of Physics for its warm hospitality. LG acknowledges support from the Australian Government through the Australian Research Council Laureate Fellowship grant FL180100168.

%
%
%

\bibliographystyle{JHEP}
\bibliography{BHDrifts}


\end{document}